%% file: main.tex
\documentclass[letterpaper,12pt]{article}

\usepackage{amssymb}
\usepackage{amsthm}
\usepackage{newtxmath}
\usepackage{newtxtext} 
\usepackage{microtype} 
\usepackage{setspace}
\usepackage{xcolor}
\usepackage{graphicx}
\usepackage{booktabs}
\usepackage{orcidlink}
\usepackage{placeins} 
\usepackage{hyperref}
\hypersetup{hidelinks,colorlinks=false}
\usepackage{pdfpages}
\usepackage{makecell} 
\usepackage{array,booktabs,tabularx}
\usepackage{tikz} 
\usetikzlibrary{positioning}
\usepackage{fontspec}
\newcommand*\circled[1]{\tikz[font=\sffamily\scriptsize,baseline=(char.base)]{
            \node[shape=circle,draw,inner sep=1pt] (char) {#1};}}
\usepackage{tabto}
\usepackage[top=1in,bottom=1in,left=1in,right=1in]{geometry}
\usepackage[font={small}]{caption}
\usepackage{multirow}
\usepackage{url}
\usepackage{anyfontsize}

\usepackage{abstract} 

\usepackage{authblk} 
\usepackage[square,numbers,sort&compress]{natbib}

\begin{document}

\title{
  \noindent
  \vspace{-36pt}\begin{flushright}{\large{PITT-PACC-2602-v1}}\end{flushright}\vspace{36pt}
  \noindent
  \textbf{
    Memristive tabular variational autoencoder for compression of analog data in high energy physics 
  }
}
\vspace{9pt}

\author[1\,\orcidlink{0000-0002-8508-8405}]{R.~Gupta}
\author[1\,\orcidlink{0009-0005-5557-7231}]{Y.~Elangovan}
\author[1\,\orcidlink{0000-0001-7834-328X}]{T.~M.~Hong\thanks{\href{mailto:tmhong@pitt.edu}{tmhong@pitt.edu}}}
\author[2\,\orcidlink{0000-0001-5091-3674}]{J.~Ignowski}
\author[2\,\orcidlink{0000-0002-6516-2700}]{J.~Moon}
\author[2\,\orcidlink{0000-0001-5260-4111}]{A.~Natarajan}
\author[1,3\,\orcidlink{0000-0002-3878-5873}]{S.~T.~Roche}
\author[2\,\orcidlink{0000-0002-9646-0970}]{L.~Buonanno\thanks{\href{mailto:luca.buonanno@hpe.com}{luca.buonanno@hpe.com}}}

\affil[1]{\fontsize{13pt}{14pt}\selectfont Pittsburgh Particle Physics, Astrophysics, and Cosmology Center,\qquad\qquad\qquad Department of Physics and Astronomy, University of Pittsburgh, Pittsburgh, PA,~USA\vspace{4pt}}
\affil[2]{\fontsize{13pt}{14pt}\selectfont HPE Labs, Fort Collins, CO,~USA\vspace{4pt}}
\affil[3]{\fontsize{13pt}{14pt}\selectfont School of Medicine, Saint Louis University, St. Louis, MO, USA\vspace{4pt}}

\date{\today}

\maketitle
\begin{abstract}
\noindent
We present an implementation of edge AI to compress data on an in-memory analog content-addressable memory (ACAM) device. A variational autoencoder is trained on a simulated sample of energy measurements from incident high-energy electrons on a generic three-layer scintillator-based calorimeter. The encoding part is distilled into tabular format by regressing the latent space variables using decision trees, which is then programmed on a memristor-based ACAM. In real-time, the ACAM compresses $48$ continuously valued incoming energies measured by the calorimeter sensors into the latent space, achieving a compression factor of 12x, which is transmitted off-detector for decompression. The performance result of the ACAM, obtained using the Structural Simulation Toolkit, the SST open source framework, gives a latency value of $24\,$ns and a throughput of $330\times10^6\,$compressions per second, i.e., $3\,$ns between successive inputs, and an average energy consumption of $4.1\,$nJ per compression.
\end{abstract}
\vspace{36pt}
    Trigger concepts and systems (hardware and software).
\vfill

\section*{Introduction}

\begin{figure}[b!]
  \centering
  \includegraphics[width=\textwidth]{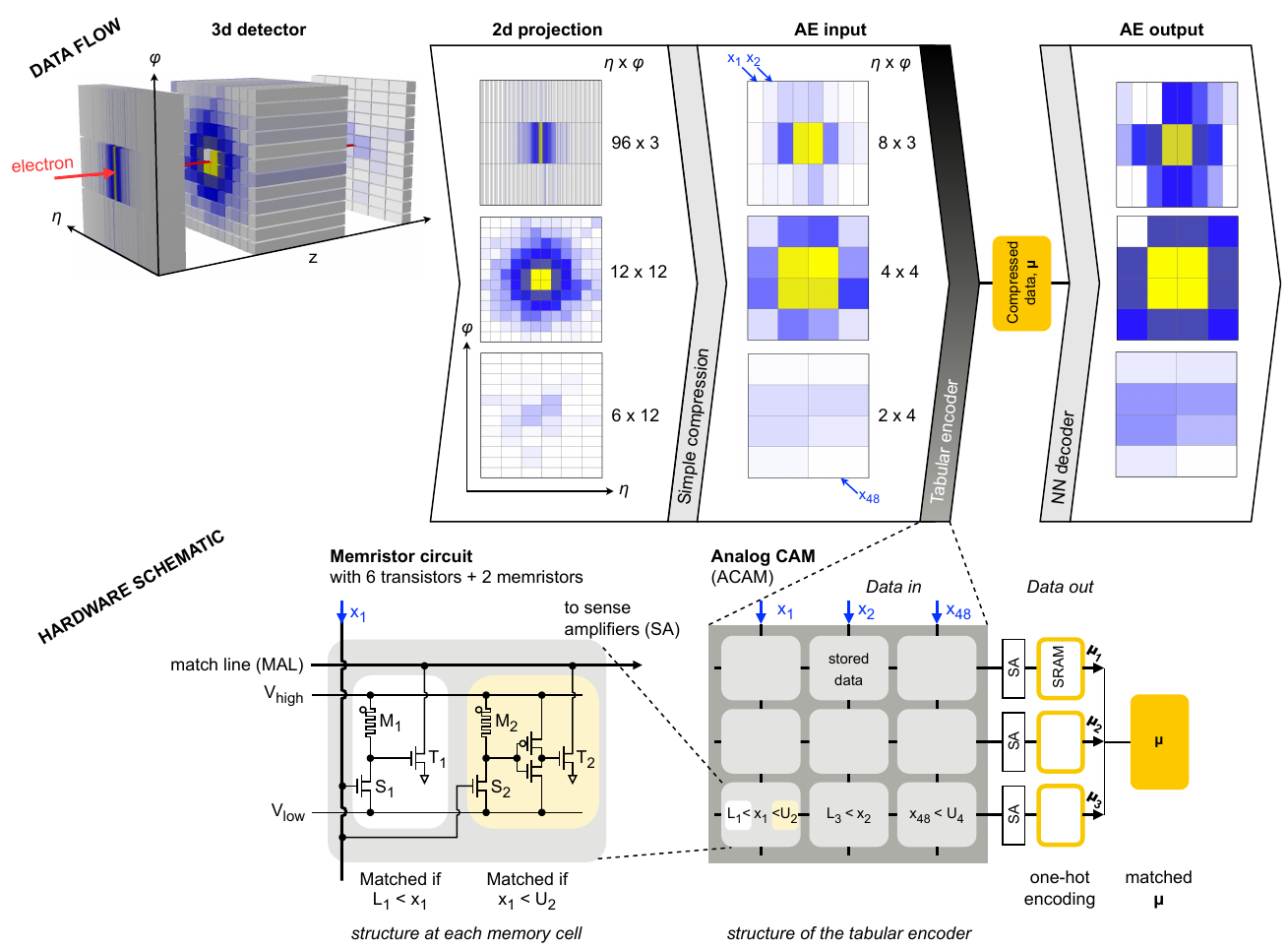}
  \caption{
    Schematic of an autoencoder for data compression that is distilled into tabular format. Top: The dataflow starts with an incident electron, here with $83\,$GeV of energy, traversing a three-layer calorimeter. The energy deposits are projected onto the transverse planes, which are then simplified by grouping energies of nearby sensor elements, which serves as input to the tabular AE. Bottom: Close-up of a memristor-based analog content-addressable memory shows the crossbar structure of the input data ($\mathbf{x}$) crossing the match line to read-out into static RAM. Further close-up at each crossbar shows the memristor circuit architecture to produce a binary output. The latent data is transmitted and decompressed.
  }
  \label{fig:io}
\end{figure}

The growth of data in high energy physics is exploding, mirroring the trend in industry \cite{BigData}. We describe the challenges, and our solution, of the data acquisition systems for proposed experiments at a future lepton collider. For instance, the electron-electron version of Future Circular Collider (FCC) called FCC-ee \cite{FCC:2018evy} or the Muon Collider called $\mu$C \cite{Long:2020wfp,Accettura:2023ked} may need a streaming data acquisition system to readout the collisions. For these and other such large-scale systems, we consider an autoencoder (AE) from artificial intelligence (AI) that is distilled and implemented on a power-efficient memristive analog memory device to compress analog data near the front-end. We consider a generic scenario of compressing energy deposits measured by a three-layer calorimeter detector system in a streaming readout setup. The setup and dataflow are shown on the top row of \figurename\:\ref{fig:io}; the bottom row shows the hardware implementation, which is detailed later in \emph{Methods}.

Lepton collisions---in contrast to the proton collisions or collisions involving heavy ions at, e.g., the Large Hadron Collider (LHC) at CERN---produce partonic-level interactions that yield a near-unity fraction of the collision rate to save offline for further study. In terms of dataflow, it is not inconceivable to have a billion channels operating at tens of kHz of the collision rate \cite{Apresyan:2023frr}. Regardless of whether it is necessary to save the full resolution data, it is desirable to compress data to reduce storage requirements if it is possible to maintain key physics attributes of the collision. We focus on the compression scheme that uses variational AE (VAE). In the literature many related ideas exist. For instance, implementations of VAE on field programmable gate arrays (FPGA) are focused on its use as anomaly detector  \cite{Govorkova:2021utb}, where a score is computed by comparing the input quantities and the decompressed quantities. Other FPGA-based VAE implementations focus on the simplification of the neural networks by putting a penalty on complexity \cite{CMS-DP-2023-086,Cohen:2938881} and distillation of the model using decision tree regressors \cite{Gupta:2942542}. There are other non-neural approaches such as using density estimates in the input space that naturally result in decision tree models \cite{Cao_2025,Roche:2023int}. Lastly, an implementation on application-specific integrated circuit (ASIC) presents a method to compress data on front-end \cite{DiGuglielmo:2021ide}. While innovations on this front continue, we note that all of these approaches may be susceptible to scaling limitations that may be overcome with in-memory computing (IMC) that is used in our approach. A brief background on IMC \emph{on the edge} is presented before discussing our setup.

In general, the performance of AI inference is increasingly constrained by the memory wall~\cite{gholami2024aimemorywall}. For example, in deep learning workloads, moving weights and activations between separate computing and memory units leads to increased latency and to limited scalability and integration, and to low energy efficiency~\cite{8114708}. This bottleneck becomes a significant limitation in data-intensive applications, like the scenario mentioned above \cite{Shanahan:2022ifi}. The IMC paradigm~\cite{ielmini_-memory_2018, mutlu2022modern, pedretti_tree-based_2021, khaddam2022hermes, garzon20234} moves away from conventional von\,Neumann computing architectures and addresses the memory wall issue by co-locating storage and computation. New integrated circuit topologies based on emerging memory technologies, such as resistive RAM (ReRAM or RRAM) and phase-change memory (PCM), or on classic memory cells, such as static RAM (SRAM), enable the parallel execution of arithmetic operations within memory. A successful example of this paradigm are the ReRAM crossbar arrays \cite{ankit2019puma, shafiee2016isaac}, which are used to perform analog dot products in situ via Ohm’s and Kirchhoff’s laws, reducing data movement and the cost per multiply-accumulate (MAC) unit. Using the IMC framework, content-addressable memory (CAM) can be built using analog components.  Mixed-signal IMC chips based on phase-change memory have demonstrated higher compute density and energy efficiency than contemporary GPU chips on targeted inference workloads~\cite{le202364, ambrogio2023analog}, while high Technology Readiness Level (TRL) digital SRAM compute-in-memory macros achieve peak energy efficiencies up to 50x those of GPU chips~\cite{dong202015}. The acceleration of small ML models at the edge for high energy physics has been demonstrated at high TRL on FPGA platforms, using on-chip URAM and BRAM and following the dataflow paradigm \cite{Roche:2023int, arora2025compute}. However, this solution is constrained by the limited on-chip memory at $\mathcal{O}(10)$ MB available for storing model parameters and weights. For our case, IMC may offer some key advantages beyond FPGA-based implementations: its weight-stationary execution model enables effectively constant-time, array-level operations, which is particularly valuable when processing high-rate collision data under tight timing constraints. Moreover, beyond crossbar arrays, our decision tree-distilled AI models naturally map to the specialized IMC memory arrays called analog content-addressable memory (ACAM).

ACAM devices can accelerate tree-based AI inference by mapping each root-to-leaf decision path to a row-and-column tabular setup: a row for each CAM module for comparisons, and a column for each feature, i.e., input variable \cite{Li2020ACAM, Yin2024FeACAM_DRF, pedretti_tree-based_2021, Halawani2021RRAM_CAM, Liu2024FeFET_ACAM}. This is well matched to the structure from the output of the model training step used in the paper, which is a set of parallel decision paths that represent regression trees \cite{Carlson:2022dgb, Serhiayenka:2024han}. A single parallel analog range-compare across all features activates the match line (MAL). The MAL directly gates a nearby SRAM word containing the leaf value.
The setup is effective because comparisons occur \textit{in memory} with lightweight periphery and inherent row-level parallelism, so latency and throughput depend mainly on the number of features and classes rather than the number of trees or their depth, avoiding load imbalance and repeated data movement on conventional platforms. Quantitatively, a simulator framework for tree-based ML accelerator based on ACAM called X-TIME reports a latency value of about $100$\,ns, throughput improvements up to $120$x, and energy-efficiency gains up to $150$x, substantially outperforming conventional, non-analog CAM hardware \cite{pedretti2024x}.
The ACAM crossbar structure and the MAL connection to SRAM is shown on the bottom row of \figurename\:\ref{fig:io}.

In this paper, we present a novel combination of the above-mentioned technologies implemented on ACAM: neural network-based VAE whose encoding part is distilled using a boosted decision tree (BDT) regressor that is converted into a tabular format for the ACAM. The procedure to create the encoder is done in four steps: AI training, model distillation, model tabularization, and model deployment on hardware. \circled{1} VAE is trained on simulated data of electromagnetic calorimeter showers to learn a compact latent representation that preserves both the longitudinal and transverse structure of the cascade of electromagnetic interactions \cite{ecal_mendeley}. The encoder maps each shower into a lower-dimensional latent space for an efficient and high-fidelity compression. \circled{2} In order to achieve hardware-friendly deployment at the detector front-end, the encoder is distilled with BDT trained to regress the latent variables using the same inputs as for the VAE \cite{Carlson:2022dgb}. \circled{3} After distillation, the model is tabularized by parallelizing the decision paths of the decision trees \cite{Carlson:2022dgb,Serhiayenka:2024han}. \circled{4} This tabular representation of the VAE encoder is deployed via root-to-leaf on the memristor-based ACAM, which performs the inference directly in memory using analog range-compare operations \cite{Graves2020IMC_CAM}. We refer the model after step \circled{4} as the \emph{tabular variational autoencoder}. Diagrammatically, the workflow is
\begingroup
\sffamily
\begin{equation}
\text{
  \footnotesize
  \begin{tikzpicture}
    \node(formula){Simulated data\ $\xrightarrow[\text{Train}]{\circled{1}}$ VAE};
    \node(solution1) [above right =-1.0em and 1em of formula]{
    Encoder
    $\xrightarrow[\text{Distill}]{\circled{2}}$
    Regressed BDT
    $\xrightarrow[\text{Tabularize}]{\circled{3}}$
    Parallel Paths
    $\xrightarrow[\text{Root-to-Leaf}]{\circled{4}}$
    ACAM
    };
    \node(solution2) [below right =-0.5em and 1em of formula]{
    Decoder
    $\xrightarrow[\text{\,As is\ }]{}$
    User hardware
    };
    \draw [->] (formula.east) to [out=0, in=180] (solution1.west);
    \draw [->] (formula.east) to [out=0, in=180] (solution2.west);
  \end{tikzpicture}
}
\label{eqn:training}
\end{equation}
\endgroup
\noindent
After the AI deployment, the dataflow of real-time processing is
\begingroup
\sffamily
\begin{equation}
\text{
  \footnotesize
  \begin{tikzpicture}
    \node(formula){Real-time analog data on front-end electronics\ 
    $\xrightarrow[\text{Tabular ACAM}]{\circled{i}}$
    Compressed digital data (near sensor)
    };
    \node(solution1) [below=0.5em and 1em of formula]{
    Decompressed digital data in DAQ system
    $\xleftarrow[\text{VAE Decoder}]{\circled{ii}}$
    Compressed digital data (away from sensor)
    };
    \draw [->] (formula.east) to [out=0, in=0] (solution1.east);
  \end{tikzpicture}
}
\label{eqn:inference}
\end{equation}
\endgroup
\noindent
We focus on $\circled{i}$ in this paper; note that the decoder in $\circled{ii}$ retains its original form. 

The outline of the rest of the paper is as follows: \textit{Results} present the physics algorithm performance and the ACAM hardware performance. Comparisons are also made to an equivalent setup on FPGA. \textit{Discussion} summarizes the limitations of our study and possible future directions. \textit{Methods} detail the neural network architecture, the training setup, the model distillation to decision trees. Details of the hardware simulation are also given.

\section*{Results}

We present the results of a neural compression pipeline that integrates deep learning with analog in-memory computing for real-time tabular data compression in high energy physics. Two results are presented: physics performance and ACAM performance. The former compares the input before \circled{1} and the output before \circled{3} in the workflow defined in (\ref{eqn:training}) at the end of the \textit{Introduction}. The latter refers to $\circled{i}$ in the dataflow defined in (\ref{eqn:inference}). Additional details, such as the comparisons of steps after \circled{3} and the breakdown of the substeps within $\circled{i}$, are given later in \textit{Methods}.

The experimental setup uses an electromagnetic calorimeter (ECAL), shown in \figurename\:\ref{fig:io}, that is simplified into 48 energy deposits in three layers by grouping nearby cells. Using widely available software, VAE model parameters are determined by training on a sample of simulated energy deposits from incident electrons; see \emph{Data availability}. The model is characterized by a set of four latent space variables $\boldsymbol{\mu}$. Each latent component of $\boldsymbol{\mu}$ is independently regressed to $\boldsymbol{\hat{\mu}}$, denoted with a circumflex to represent its estimate, using the 48-dimensional input vector $\mathbf{x}$ for the energy deposits. The main computation takes place within a core composed of an ACAM, SRAM, and an adder. Some implementations proposed in literature also include a \textit{Multiple Match Resolver} in the core \cite{pedretti2024x}. The ACAM stores the tabular version by mapping each root-to-leaf path in the tree to a row, such that the search operation generates a match vector indicating the indices of matched rows. Consequently, the number of ACAM rows and columns determines how many root-to-leaf paths can be stored and how many input features can be applied, respectively. These leaf values are accumulated by the adder, and the core’s result is then sent to the output communication. A multi-level H-Tree network-on-chip is employed in the output communication to efficiently aggregate results from multiple cores along the result retrieval path. The details of the schematic in the bottom row of \figurename\:\ref{fig:io} are shown later in \figurename\:\ref{fig:Simulation_diagram}.

\subsubsection*{Physics algorithm performance}

An on-detector compression scheme should preserve the characteristic variables as well as the image characteristics of the electromagnetic shower resulting from the incident electron. The physics content, in this case, is the energy development in the longitudinal $z$ direction, the transverse shape in the $\eta\times\phi$ plane, and the overall energy scale of the shower, and defines the following observables: total energy ($E_\textrm{tot}$), layer energies $(E_\ell)$ and their fraction with respect to $E_\textrm{tot}$ $(f_\ell=E_\ell/E_\textrm{tot})$ labeled by $\ell$, shower depth ($s_d$), and lateral widths ($\sigma_\ell$). A table of definitions are given in \emph{Supplementary Materials}. Each observable is considered in three ways:
\begin{itemize}
  \setlength\itemsep{0em}
  \item No compression (Original): Observables are formed using the 48 input energies.
  \item VAE encoding: Observables are formed after compressing the inputs using the VAE encoder, the one between \circled{1} and \circled{2} in (\ref{eqn:training}), then decompressing the latent $\boldsymbol{\mu}$.
  \item BDT encoding: Observables are formed after compressing the inputs using the distilled BDT encoder, the one between \circled{2} and \circled{3} in (\ref{eqn:training}), then decompressing the regressed $\boldsymbol{\hat{\mu}}$.
\end{itemize}

\begin{figure}[b!]
  \centering
  \includegraphics[width=1\textwidth]{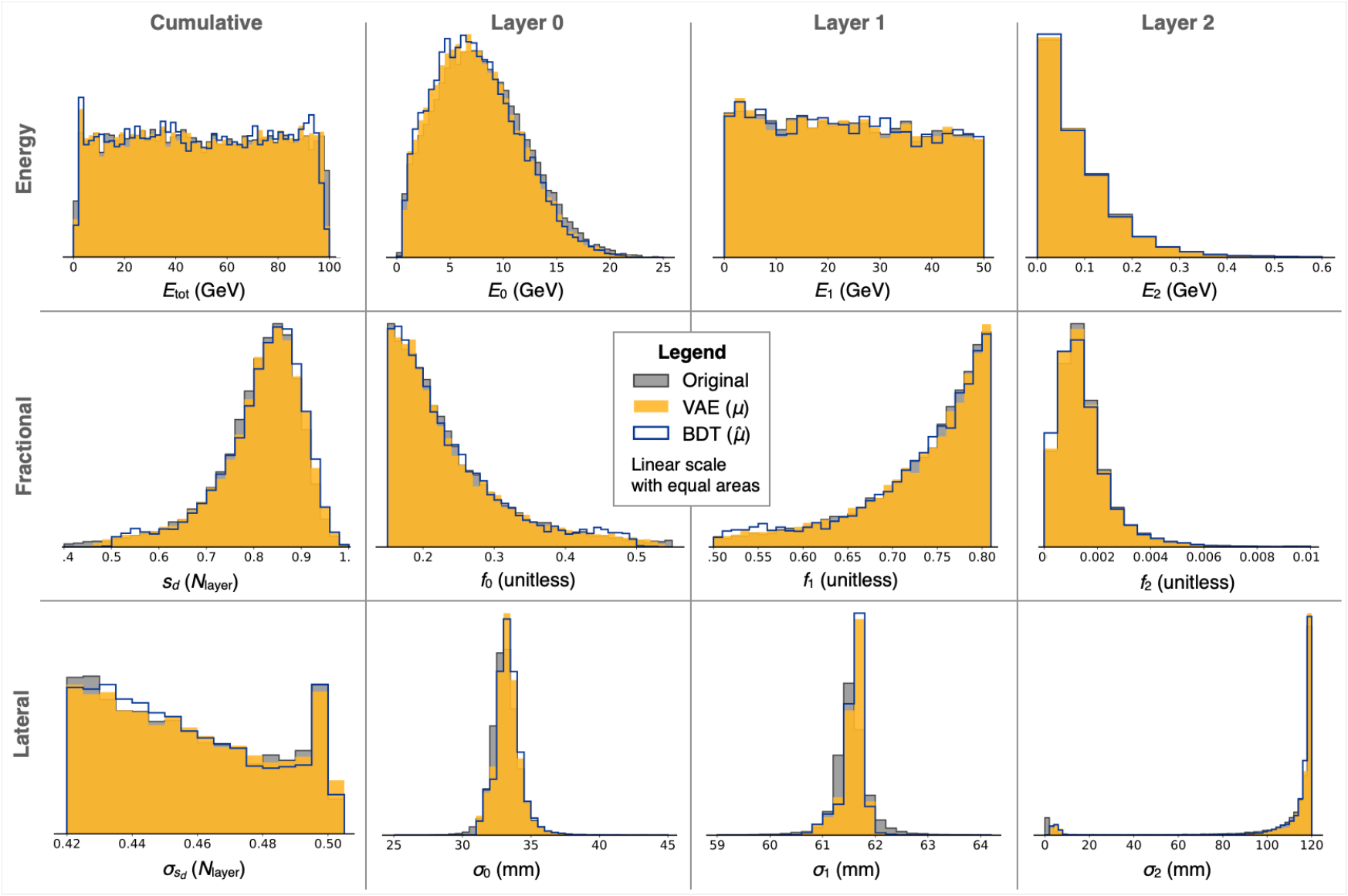}
  \caption{
    Physics observables before and after compression for the original electromagnetic showers (grey), the VAE encoder ($z=\mu$, gold), the BDT–regressed encoder ($z=\hat{\mu}$, blue line). See text.}
  \label{fig:physics_summary}
\end{figure}

A simulated sample of electrons with initial energies that are uniformly distributed up to $100\,\mathrm{GeV}$ is used to train and evaluate the AI model. The distributions of the three sets of observables for the above-mentioned list of variables are shown in \figurename\:\ref{fig:physics_summary}. After compression and decompression, $E_{\mathrm{tot}}$ and the three $E_\ell$ are well reproduced across the full energy range, with small shape differences. The corresponding fractions $f_\ell$ and the shower-depth estimators $s_d$ and $\sigma_{s_d}$ also show good agreement between the Original and the two reconstructed distributions, indicating that the longitudinal development of the shower in the $z$ direction is largely preserved. For $\sigma_{s_d}$, the peak near $0.5$ is expected since electromagnetic showers deposit negligible energy in the final layer and are, therefore, effectively shared between the first two layers. The lateral width observables, which probe higher-order transverse spatial moments of the shower development, are also reproduced with good fidelity. Small residual differences remain, visibly in the tails of the $\sigma_1$ distributions. These tails are driven by rare showers with atypical low-energy peripheral deposits and are, therefore, more sensitive to small distortions than the total energy or the longitudinal observables. Importantly, the VAE- and BDT-based reconstructions are nearly indistinguishable in all observables, demonstrating that the distillation step does not introduce an additional degradation beyond the VAE representation itself. Quantitatively, the two-sample Kolmogorov-Smirnov distance demonstrates agreement at the few percent level between all but one of the distributions; the one exception being $\sigma_1$ at $0.2$ whose tail was noted above.

The preservation of shower images at the cell level is quantified using the $L_1$ and $L_2$ metrics between the original and reconstructed cell-energy vectors, respectively, normalized to the latter. All values are found to be small at around $0.07$. These results confirm that the BDT-regression reproduce the VAE with only a small loss in image-level fidelity. A pedestrian example of image reconstruction is given using color photos of Belgian street signs in \emph{Supplementary Materials}.

\subsubsection*{ACAM hardware performance}

We report latency and throughput results for varying level of threshold precision from $4$ to $32$ bits in the encoder given by the simulation of the ACAM-based architecture. The ACAM structure varies depending on the needs of the BDT regressor. For 4-bit thresholds, the sensor’s analog output can be directly fed into the cores with minimal latency; in this case, the analog values propagating along the CAM columns are compared, at each row, with the value stored in a single memory cell by the two memristors, by $M_1$ and $M_2$ in \figurename\:\ref{fig:io}. For higher-precision thresholds,  analog-to-digital conversion (ADC) is performed to separate the value into 4-bit groups, then its converse (DAC) is performed to feed the analog architecture. For example, for 8-bits thresholds the input is digitized into 8-bits, split into two 4-bit groups, then sent to ACAM modules.

\begin{figure}[b!]
  \centering
  \includegraphics[width=1.0\textwidth]{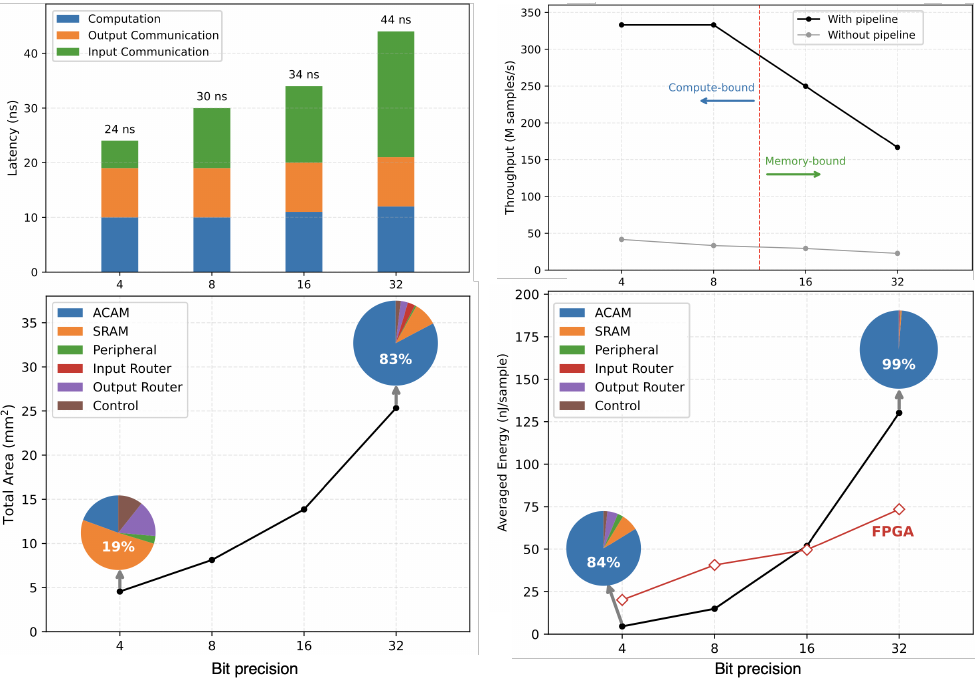}
  \caption{
    Latency (top-left), throughput (top-right), total area (bottom-left) and energy per compression (bottom-right) of the encoding part simulated on the ACAM-based architecture. \emph{Computation} latency at about 10\,ns is due to the ACAM alone. See \emph{Discussion} for the comparison to FPGA in the bottom-right plot.
  }
  \label{fig:acam_performance}
\end{figure}

The total latency is the lowest for $4$-bit precision at $24\,$ns as expected. As seen in the breakdown in the top-left plot of \figurename\:\ref{fig:acam_performance}, the latency component associated with computation is independent of bit precision at $10\,$ns. This is also true for output communication. However, the component associated input communication grows with bit precision. This is because input communication, given a fixed memory bandwidth, scales directly with the total size of input data, and is determined by the product of the number of features and the feature bit-width.

The throughput is the inverse of latency for dataflow that is not pipelined. As seen in the top-right plot of \figurename\:\ref{fig:acam_performance}, the non-pipelined throughput is relatively flat around $40\,$M compressions per second.  By introducing pipelining into the dataflow, the throughput can be substantially improved up to eightfold to a peak value of $330\,$M compressions per second. The throughput gradually decreases with precision at 8 bits linearly down to around $150\,$M for 32 bits. This trend reflects the well-known trade-off between compute-bound and memory-bound regimes. When the data size is sufficiently small such that data transfer completes faster than computation, performance is limited by computation. Conversely, as the data size grows, data transfer becomes the dominant performance bottleneck.

Lastly,  the efficiency of implementation is considered by presenting the total area of the implementation and the average energy consumed per compression shown in the bottom two plots of \figurename\:\ref{fig:acam_performance}. As described later in \emph{Methods}, high-precision ACAM computation is performed recursively and require multiple stages of ACAM blocks. Consequently, both the area and the energy approximately increase in proportion to the precision, although the ACAM inherently exhibits data-dependent energy consumption characteristics. Similar to the latency and throughput analysis, the best area efficiency (2.1 mm$^2$) and energy efficiency (4.1 nJ per compression) are achieved at 4 bits. The pie charts inside the plots gives the breakdown for 4- and 32-bit cases. The ACAM occupies the dominant share of both total area and energy consumption for the cases higher than 4 bits. This dominance arises because the entire ACAM array remains active during each search operation. The overall energy can be reduced by partitioning the ACAM array into smaller sub-arrays and activating only the sub-arrays corresponding to previously matched stages, or by early termination of the search when a mismatch is detected in the most significant bits of the recursive computation.

\section*{Discussion}

A summary is given followed by a discussion of four topics. First, a comparison is made to results of deploying the distilled AI on FPGA. Second, analog CAM for tree-based ML is reviewed. Third, IMC designs in high energy physics is considered. Lastly, anomaly detection at colliders is discussed.

\subsubsection*{Summary}

In this paper, we present an end-to-end compression pipeline that combines a VAE trained on ECAL showers with a distilled, tree-based surrogate implemented on analog content-addressable memory. The VAE learns a 4d latent representation of $48$ rebinned calorimeter features, achieving an effective $12$x compression while preserving the longitudinal and transverse structure of ECAL showers. Using physically motivated observables such as total energy, layer-wise fractions, shower depth, and lateral width, we find that reconstructions based on the BDT-regressed latent vectors are statistically indistinguishable from those obtained with the neural encoder, indicating that the distillation and tabularization steps do not introduce additional loss of physics information.

On the hardware side, the ACAM-based implementation delivers ns-scale latency and few-nJ-per-compression for energy consumption in the most efficient precision regime. The data throughput achieves hundreds million compressions per second in the pipelined design. This behavior reflects the intrinsic advantages of in-memory, row-parallel range compares over the conventional von\,Neumann designs.

\subsubsection*{FPGA implementation}

For a digital baseline, we implement the distilled BDT encoder on an Versal FPGA from AMD model VP1802 using the Vivado software package version 2024.2. The design is deployed and tested on a VPK180 evaluation board using \textit{Virtual Input/Output} cores to supply test vectors and capture results on hardware. All designs are evaluated at $300$\,MHz, which is the highest frequency at which timing closure was achieved. The models use on-chip LUT RAM to store the input features, a state-machine controller to sequence read-compute-write operations, and an \textit{Integrated Logic Analyzer} to measure cycle-accurate latency on hardware. Power is estimated using post-implementation activity data.

Results are given for latency, energy consumption, and resource utilization. Latency is considered across the four bit precisions. The FPGA design exhibits a fixed pipeline depth of $13$ clock cycles, corresponding to a latency value of $43\,$ns. This value is about twice higher than for the best ACAM result at 4 bits and comparable to the worst at 32 bits. This precision-independent behavior reflects the deeply pipelined digital comparator and accumulation stages. In contrast, the ACAM architecture shows a precision-dependent latency profile: the computation and output communication remain nearly constant, while the input communication increases proportionally to the number of transmitted bits. Energy  consumption on the FPGA, shown in \tablename\:\ref{tab:fpga_results}, ranges from 20\,nJ per compression for 4\,bits to 74\,nJ per compression for 32\,bits. The ACAM energy also increases with precision, but the absolute values differ substantially: at 4\,bits the ACAM requires a much lower value at 4\,nJ per compression sample, a fivefold advantage. However, for higher precisions the recursive multi-stage ACAM computation and static biasing raise the total energy above the comparable FPGA result. Resource utilization on the FPGA scales strongly with input precision. The LUT count rises from 146\,k for 4\,bits to 365k for 32\,bits, with a corresponding increase in slice and register usage. The number of \textsc{Lookahead8} units remains constant, reflecting the fixed structural depth of the BDT. Although a direct comparison to ACAM area is not meaningful due to fundamentally different architectures, these results highlight the expected digital scaling behavior: widening the input thresholds primarily increases comparator logic, routing, and the associated pipeline registers.

\newcolumntype{M}[1]{>{\centering\arraybackslash}m{#1}}
\newcolumntype{L}[1]{>{\raggedright\arraybackslash}m{#1}}
\begin{table}[bh!]
  \centering
  \caption{
    FPGA implementation of the encoder. Latency, resource utilization (LUT, registers, \textsc{Lutram}, Slices, \textsc{Lookahead8} units), and energy per compression are given for input widths from 4 to 32 bits.
  }
  \label{tab:fpga_results}
  \renewcommand{\arraystretch}{0.90}
  \setlength{\tabcolsep}{4.5pt}
  {\small
  \begin{tabular}{M{4.5cm} L{3.9cm} r r r r}
    \toprule
    \multirow[c]{3}{4.5cm}{\centering\textbf{Design} \textbf{parameters}}
    & Input width (no.\:bits)   & 4    & 8    & 16   & 32   \\
    & Output width (no.\:bits)  & 16   & 16   & 16   & 16   \\
    & Clock speed (MHz)         & 300  & 300  & 300  & 300  \\
    \midrule
    \multirow[c]{2}{4.5cm}{\centering\textbf{Timing}}
    & Clock ticks               & 13   & 13   & 13   & 13   \\
    & Latency (ns)              & 43   & 43   & 43   & 43   \\
    \midrule
    \multirow[c]{5}{4.5cm}{\centering\textbf{Resource} \textbf{utilization}}
    & LUT                       & 146\,k & 146\,k & 197\,k & 365\,k \\
    & Register                  & 56\,k  & 75\,k  & 80\,k  & 88\,k  \\
    & \textsc{Lutram}           & 2\,k   & 2\,k   & 2\,k   & 2\,k   \\
    & Slice                     & 12\,k  & 26\,k  & 34\,k  & 63\,k  \\
    & \textsc{Lookahead8}       & 2\,k   & 2\,k   & 2\,k   & 2\,k   \\
    \midrule
    \parbox[c]{4.5cm}{\centering\textbf{Energy} \textbf{consumption}}
    & \vspace{2pt} Energy (nJ)  &  20   &  41   & 50   & 74   \\
    \bottomrule
  \end{tabular}
  }
\end{table}

An additional distinction arises at the detector front-end. For low threshold precisions at 4\,bits the ACAM can operate directly on the analog detector output, removing the need for an ADC before the encoder. For higher precisions, a small ADC is required to generate the multi-bit representations used in the recursive ACAM computation. The FPGA implementation, in contrast, always operates on fully digital inputs and, therefore, always requires an ADC at the front-end. Overall, the comparison indicates that ACAM and FPGA architectures excel in different operating regimes. The ACAM provides substantially lower energy per compression and reduced front-end complexity in the low-precision regime relevant to threshold-based BDT inference, while the FPGA offers precision-independent latency and predictable digital scaling at higher bit widths. We believe that improved optimization of the engineering of the higher precision setup is possible, but is beyond the scope of this paper.

\subsubsection*{Analog CAM for tree-based ML}

The advent of IMC architectures aims to mitigate the memory-wall in von\,Neumann machines by performing computation where data reside. IMC crossbar arrays based on non-volatile memristors enable energy-efficient matrix multiplication and have been used to accelerate machine learning, analog signal processing and scientific computing \cite{shafiee2016isaac}.  However, these crossbars are optimized for matrix-vector multiplications (MVM) and do not natively support other data intensive operations. Parallel search is a fundamental operation in databases, pattern matching and machine learning inference, yet conventional parallel search architectures such as CAM remain largely digital for off-the-shelf products, and suffer from high power consumption and limited functionality. Recent work seeks to extend IMC concepts to CAM by exploiting the analog programmability of emerging memory devices \cite{Li2020ACAM,Liu2024FeFET_ACAM}.

Tree-based  models are widely used because they train quickly, perform well on small data sets, and provide interpretability. Yet their inference is challenging to optimize in von\,Neumann architectures due to irregular memory access patterns \cite{pedretti2024x}. Pedretti et al.\ mapped each root-to-leaf path of a decision tree onto a row of an ACAM, encoding decision boundaries as analog intervals. The search operation across the ACAM array simultaneously evaluates all possible paths, enabling few-cycle inference and dramatically improving throughput \cite{pedretti_tree-based_2021}.  Similarly, Yin et al.\ showed that a ferroelectric ACAM consisting of two ferroelectric field-effect transistors (FeFET) can implement the branch-split operations in deep random forest accelerators, with energy-efficient associative searches that store decision boundaries as FeFET polarization states \cite{Yin2024FeACAM_DRF}. Their experiments indicated that the FeFET-based ACAM accelerator achieves orders-of-magnitude improvements in energy and latency compared with digital implementations. Liu et al.\ experimentally demonstrated an ACAM cell using complementary FeFET that stores more than forty distinct match windows and performs parallel search without requiring external analog converters \cite{Liu2024FeFET_ACAM}. The FeFET ACAM uses only two transistors per cell, merging computation and analog-to-digital conversion to achieve higher density and even lower power than ReRAM-based designs. Lastly, Halawani et al.\ combined ReRAM-based CAM with a time-domain analog adder to compute similarity scores in hyperdimensional computing, reducing area and energy by more than an order of magnitude relative to digital implementations while using a winner-take-all circuit to select the best match \cite{Halawani2021RRAM_CAM}. These works illustrate how ACAM can serve as computational primitives for non-linear models beyond MVM-centric neural networks.

Although ReRAM is still an emerging technology, multiple chip-level demonstrations have shown that it can deliver energy- and area-efficient AI acceleration without degrading the model accuracy  \cite{wan2022compute,xue2021cmos,hung2021four,chen2019cmos,yao2020fully}. The ReRAM-based accelerator for tree-based ML proposed here are not specific to ReRAM, and could be realized with alternative memory technologies---such as PCM \cite{burr2016recent}, FeFET \cite{jerry2017ferroelectric}, or SRAM-based macros \cite{jaiswal20198t}---with only minor performance differences. We propose and profile the performances of the ReRAM-based circuit owing to its large resistive window (a larger dynamic range of the resistances implies more information density), low-power operation (higher resistances dissipate less current), multilevel programmability, CMOS-compatible low switching voltage \cite{mannocci2023memory}, straightforward fabrication of high-density memory macros with crossbar arrays, and our extensive prior experience with the technology \cite{sheng_lowconductance_2019}.

\subsubsection*{IMC in high energy physics}

Other analog IMC accelerators are proposed,
in the context of radiation tolerance and calibration under the extreme environmental conditions of particle physics experiments, as a candidate technology for next-generation front-end electronics \cite{albrecht2025summary,kosters2023benchmarking}. For instance, the accelerator proposed by K\"osters et al., which employs an artificial neural network for anomaly detection at the LHC \cite{kosters2023benchmarking}, reports a gain of over three orders of magnitude higher energy efficiency and 20x faster throughput compared to CPU and GPU implementations. These improvements stem primarily from efficient matrix–vector multiplications within memory crossbar arrays and the use of a pipelined dataflow. Under the same precision setting of 8-bit input and threshold, our approach achieves 14\,nJ per compression and 333\,M compression per second, corresponding to a 10x improvement in energy efficiency and a 16x increase in throughput over the analog IMC design that reports 150\,nJ per compression and 20\,M compression per second. The gain is likely attributable to the parallel search operations of the ACAM and the finer-grained pipelined dataflow employed in our design.

Other implementations of tree-based models exist, such as SRAM-based IMC ASIC~\cite{kang2018ASIC}. Considering a comparable setup, our design achieves 1.4x higher energy efficiency and three-orders-of-magnitude faster throughput. While our total energy consumption is dominated by the static power of preprogrammed ACAM, which is necessary to avoid frequent reprogramming, the SRAM-based IMC reduces redundant static energy through the highly utilized SRAM array. Consequently, energy efficiency is comparable, but the throughput is significantly enhanced in the ACAM.

Looking ahead, our strategy of neural compression, tree-based distillation, and analog in-memory execution may be applicable to other systems with different geometries. Crossbar arrays exist, and are optimized for neural network inference via MVM, the matrix-vector multiplications described above in the discussion of the ACAM. These IMC architectures have demonstrated transformative potential across diverse edge AI applications---from robotics and healthcare to aerospace systems---all of which demand real-time decision-making under tight power budgets \cite{lu2024high}. A hybrid design that combines our tree-based distillation with some of the original neural network using crossbar arrays may be advantageous for situations that require higher physics performance. Lastly, the joint task of compression and anomaly detection could provide a flexible path toward an intelligent, low-power front-end electronics at future colliders, which we discuss next.

\subsubsection*{Anomaly detection at colliders}

Other experimental environments, like the hadron-hadron version of the FCC (FCC-hh) \cite{FCC:2018vvp}, will likely require large data reduction in real-time using a trigger system. FCC-hh's trigger design can be understood by considering the LHC. Like the colliders before it, the LHC continues to increase the data-taking rate in order to collect more data in less time. This is done by improving the beam luminosity to raise the number of simultaneous interactions per bunch crossing. This phenomenon is reminiscent of multiple photographic exposures on one printed sheet and is called \emph{pileup}. Before the LHC, the Tevatron started at pileup below 1 in the 1990s and it grew six-fold by the late 2000s. Likewise, the LHC started at pileup of 10 in 2010 and it again grew six-fold. The corresponding yearly samples grew 100x in the same time period. From 2030 for about a decade, upgrades are planned for pileup of 200. After that, pileup at the FCC-hh may reach 1000 \cite{FCC:2018vvp}. FCC-hh's trigger will likely mimic the existing ones at ATLAS and CMS experiments \cite{ATLAS:2023dns,CMS:2016ngn}, which have recently deployed autoencoders, or parts thereof, as anomaly detectors, on FPGA-based systems. This follows a long build-up of AI technologies on FPGA \cite{Hong:2021snb,Carlson:2022dgb,Roche:2023int,Serhiayenka:2024han,Duarte:2018ite,Aarrestad:2021zos,Ghielmetti:2022ndm,Loncar:2020hqp,Neu:2025hst,Sun:2025orx,Sun:2025ssy,Badea:2024zoq,Borella:2024mgs,Yoo:2023lxy,Neu:2023sfh,Coccaro:2023nol,Abidi:2022ogh,Sun:2022bxx,Butter:2022lkf,Jwa:2019zlh} that led experiments to include more sophisticated designs \cite{Spolidoro_Freund_2020,Gupta:2933128,Jiang:2024vvn,Voigt:2915250,Bileska:2025jxv,Gandrakota:2024yqs,Astrand:2025sij}. Due to their autoencoder origins, we speculate that anomaly detectors may be combined along with data compression, like the one proposed in this paper.

An interesting use case may be at the $\mu$C, where complications may arise  from beam-induced background (BIB), which is caused by muons decaying in flight and their interaction with detector elements. Like the pileup above, BIB may introduce a high level of stochastic noise \cite{Holmes:2025ybv} on top of the desired muon-muon collision. In this scenario, it may be necessary to suppress BIB while simultaneously providing a steady real-time data acquisition like the FCC-ee case considered in the \textit{Introduction}. Both tasks may be possible with an autoencoder, with the latter task employing it as an anomaly detector trained on BIB data to help select non-noisy regions to compress.

\section*{Methods}

The experimental method is described in four parts. First, the data sample is presented. Second, the VAE provides the initial model of shower compression and decompression. Third, the model is distilled into decision tree regressors into tabular format. The reconstruction fidelity of the showers is considered along with the descriptions. Lastly, deployment on ACAM is presented.

\subsubsection*{Data sample}

The experimental setup considers a segmented ECAL. ECAL systems in modern collider detectors, e.g., ATLAS and CMS \cite{ATLAS:1996ab,CMS:1997ysd}, are longitudinally segmented, optimized for a different stage of the shower evolution. The first layer provides fine granularity in $\eta$ to capture the narrow electromagnetic core, while the second and third layers are progressively coarser, recording the later stages of shower propagation and energy dispersion in depth. Each layer measures the transverse energy deposited by the incident particle, forming two-dimensional maps in pseudorapidity-azimuth ($\eta$-$\phi$) space that together describe the full shower morphology. The sample used in this study is from Ref.~\cite{ecal_mendeley} that was generated for Ref.~\cite{Paganini:2017dwg}, so we briefly summarize the relevant points here. A sample of 100\,k events are produced using \textsc{Geant\,4}~\cite{GEANT4:2002zbu}, a full simulation event generator to model the behavior of particles in an ECAL. Each event considers an incident electron that deposits energy in the three layers. From these deposits, physical variables are derived, such those listed in \textit{Results}.

\begin{figure}[bt!]
  \centering
  \includegraphics[width=\textwidth]{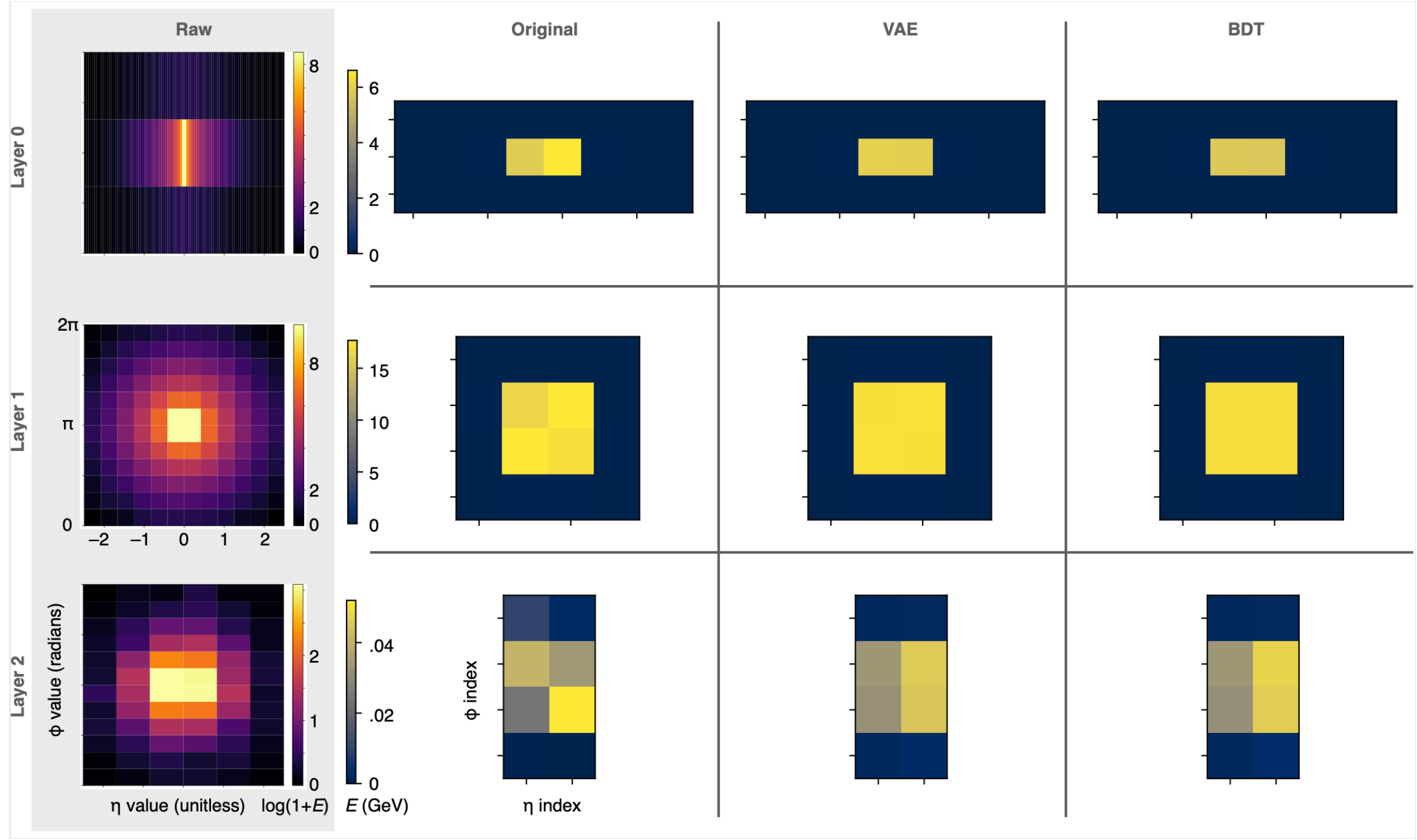}
  \caption{
    Visualization of energy deposition in $\eta$-$\phi$ for the ECAL layers in rows. Leftmost column in gray: average energy deposition from 100 simulated electron shower events with the color scale corresponding to $\log(1+E)$, where $E$ is in GeV. Right three columns show one representative shower after rebinning with color scale corresponding to linear $E$: \emph{Original} designates the input value to the VAE, \emph{VAE} is the reconstruction using the encoder mean $\mu$, and \emph{BDT} is the reconstruction using the regressed mean $\hat{\mu}$.
  }
 \label{fig:shower_image}
\end{figure}

The layers have the following segmentation: Layer~0 with $288$ cells with fine $\eta$ segmentation, $96\times3$ in $\eta\times\phi$, to resolve the electromagnetic core; Layer~1 with $144$ cells in a symmetric configuration of $12\times12$ that samples the mid-depth energy spread; and Layer~2 with $72$ cells in a coarser granularity of $6\times12$ that is optimized for the late and more diffuse part of the shower. The layers yield a total of 504 individual cells. For use in the VAE, these cell energies are preprocessed into standardized, fixed-length feature vectors that preserve the essential layer structure. The preprocessing step includes an energy-preserving spatial rebinning by summation within fixed regions, balancing compactness and physics fidelity while preserving the total deposited energy and the large-scale longitudinal energy flow. Specifically, Layer~0 is rebinned from $96\times3$ to $8\times3$, Layer~1 from $12\times12$ to $4\times4$, and Layer~2 from $6\times12$ to $2\times4$. This results in a $\mathbb{R}^{48}$ representation for the three layers. All training and quantitative evaluations are performed with energies in linear scale. The plots in the leftmost column of \figurename\:\ref{fig:shower_image} illustrates the energy deposition, averaged over 100 simulated electron showers, and shows the typical evolution of electromagnetic cascades across the ECAL layers. For visualization, the color scale is shown as $\log(1+E)$ to enhance the visibility of low-energy deposits in the shower periphery. In Layer~0, the electromagnetic core is sharply localized in $\eta$ due to fine segmentation, capturing the narrow lateral spread of the primary interaction. Layers~1 and~2 exhibit increasingly diffuse and broader energy patterns as the shower penetrates deeper into the absorber material, reflecting the combined effects of multiple scattering, energy leakage, and the coarser granularity of the later layers. This progression from compact to extended profiles encodes both the transverse and longitudinal development of the shower.

\subsubsection*{Variational autoencoder}

\begin{figure}[t!]
  \centering
  \includegraphics[width=1.0\textwidth]{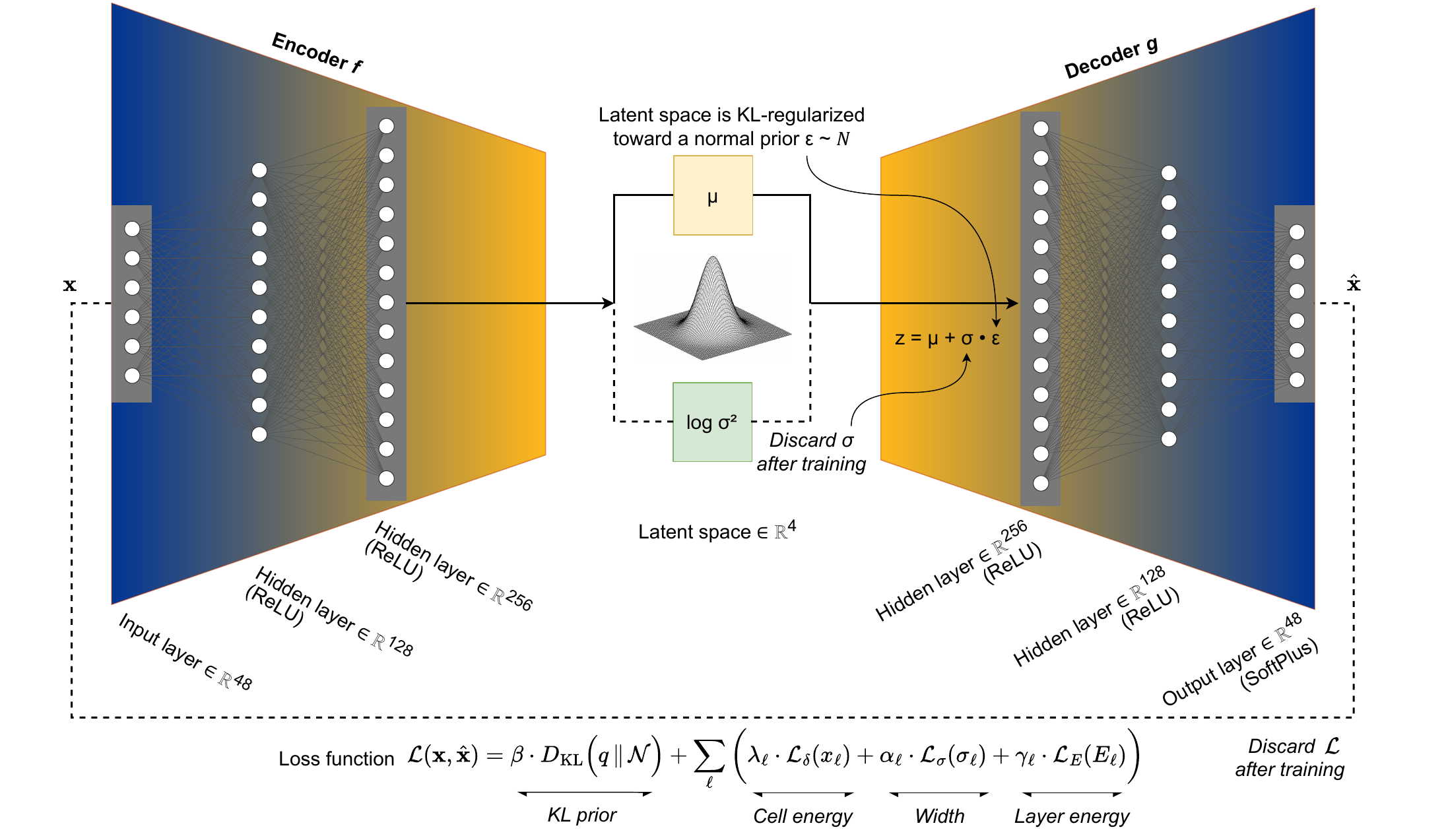}
  \caption{
    Architecture of the variational autoencoder prior to distillation. The encoder maps the calorimeter input into a latent representation parameterized by the mean and log-variance, $\mu,\log\sigma^2\in \mathbb{R}^4$. During training, latent vectors are sampled using the reparameterization trick, while inference is performed deterministically with $z=\mu$. The decoder reconstructs the shower representation with a non-negative constraint.
  }
  \label{fig:vae_architecture}
\end{figure}

The VAE provides the baseline learned model for compressing and reconstructing ECAL showers. By jointly minimizing a reconstruction objective and a Kullback-Leibler (KL) divergence regularizer, it learns smooth latent representations that capture correlations across calorimeter layers and physics-driven dependencies within each shower image. The encoder defines the target mapping for downstream distillation, while the decoder is used offline for reconstruction. The encoder $f$ that compresses the $\mathbf{x}\in\mathbb{R}^{48}$ into a latent representation $\mathbf{z}$, and the decoder $g$ takes $\mathbf{z}\in\mathbb{R}^d$ to reconstruct $\hat{\mathbf{x}}\in\mathbb{R}^{48}$, where $d$ is for the latent space:
\begin{equation}
  \mathbf{x}\xrightarrow{\textrm{encoder}\,f}\mathbf{z}\xrightarrow{\textrm{decoder}\,g}\hat{\mathbf{x}}
\end{equation}
The functions $f$ and $g$ both consist of two hidden layers (these are neural network layers, not the physical ECAL layers discussed above). The $f(\mathbf{x})$ consists of fully connected layers of 128 and 64 neurons, respectively, with ReLU activations. The latter is followed by two linear projections that produce the latent mean $\mu\in\mathbb{R}^{d}$ and log-variance $\log\sigma^2\in\mathbb{R}^{d}$, defining a diagonal Gaussian posterior $q(z|x)$. Unless otherwise stated, we use a latent dimension of $d=4$ for a twelve-fold compression factor. During training, the latent vectors are sampled via the reparameterization trick, $z=\mu+\sigma\cdot\varepsilon$, where $\varepsilon\sim\mathcal{N}(0,I)$ is drawn from a normal distribution. The $z$ is simplified for inference, where we set $z=\mu$ to ensure reproducibility and compatibility with deterministic implementations. Similar to $f$, the $g(\mathbf{z})$ consists of fully connected layers of size 64 and 128 neurons in the reverse order to produce $\hat{\mathbf{x}}$, with a SoftPlus output activation to enforce non-negative reconstructed energies. Finally, the output $\hat{\mathbf{x}}$ is reshaped into the geometries corresponding to the ECAL cells. \figurename\:\ref{fig:vae_architecture} illustrates and summarizes the VAE architecture.

An  appropriate loss function is the key to the VAE's intelligence. Ours combines a KL regularizer and a layer-weighted Huber loss at the cell level with physics-informed constraints.
\begin{equation}
  \mathcal{L}(\mathbf{x},\hat{\mathbf{x}}) = \beta\cdot D_{\mathrm{KL}}\big(q\,\|\,\mathcal{N}\big)
  \;+\; \sum_{\ell=0}^{2}
  \Big(
    \lambda_\ell\cdot\mathcal{L}_\delta(x_\ell)
    \;+\;
    \alpha_\ell\cdot \mathcal{L}_\sigma(\sigma_\ell)
    \;+\;
    \gamma_\ell\cdot \mathcal{L}_{E}(E_\ell)
  \Big)
\end{equation}
The first term is the KL divergence between the distribution of $z$, $q(z\,|\,x)$, and the normal distribution $\mathcal{N}(0,I)$. The second term is the Huber term that compares input cell energies $\mathbf{x}_\ell$ with the estimated cell energies $\hat{\mathbf{x}}_\ell$ for each ECAL layer $\ell$. Because of cell-level comparisons, this term improves robustness against rare outliers in the output image quality. The third term is the mean squared error (MSE) between the input and output transverse shower widths by matching the $\eta$-width computed from the energy-weighted $\eta$ profile. The last term is the MSE between the input total energy $E_\textrm{tot}$ and the estimated value by summing $\hat{\mathbf{x}}_\ell$ over all cells for each ECAL layer. The energy term ensures the normalization of the each layer's energy is well reproduced.

Hyperparameters are determined to resolve the difficulties we encountered on two fronts. On the one extreme, the VAE was having difficulty reconstructing the wide range of energy values that span eight orders of magnitude; see \figurename\:\ref{fig:shower_image}. To address this, we tried transforming cell energy to $\log(1+E)$ for inputs and transformed back by its inverse after the output. While this approach was successful in addressing the scale issue, the width variables were affected because it was blurring finer grained differences for each layer. The final setup that resolved both issues was take a linear scale, but to scale-up the loss function of Layer 2. The resulting values are
$\beta=0.0005$,
$\boldsymbol{\lambda}=(1,1.5,4)$,
$\delta=0.01$ for the Huber function,
$\boldsymbol{\alpha}=(0.5,0.5,5)$, and
$\boldsymbol{\gamma}=(0.005,0.005,0.03)$.

The training and testing is done on the sample using an 80-20 split, respectively. Parameters are optimized with Adam for 40 epochs with a learning rate of $10^{-3}$ and batch size of 256. Convergence is monitored through the total loss and its components, with fixed random seeds for reproducibility. A latent dimension of $d=4$ was selected after empirical studies balancing compression ratio, reconstruction fidelity, and latent smoothness. In our experience, a larger latent dimensions provided only marginal gains while smaller dimensions led to visible loss of longitudinal detail in the later layers. After convergence, the decoder reproduces both the total energy and the layer-resolved spatial patterns. One representative example is shown on the right three columns of \figurename\:\ref{fig:shower_image}.

\subsubsection*{Model distillation and tabularization}

\begin{figure}[b!]
  \centering
  \includegraphics[width=1.0\textwidth]{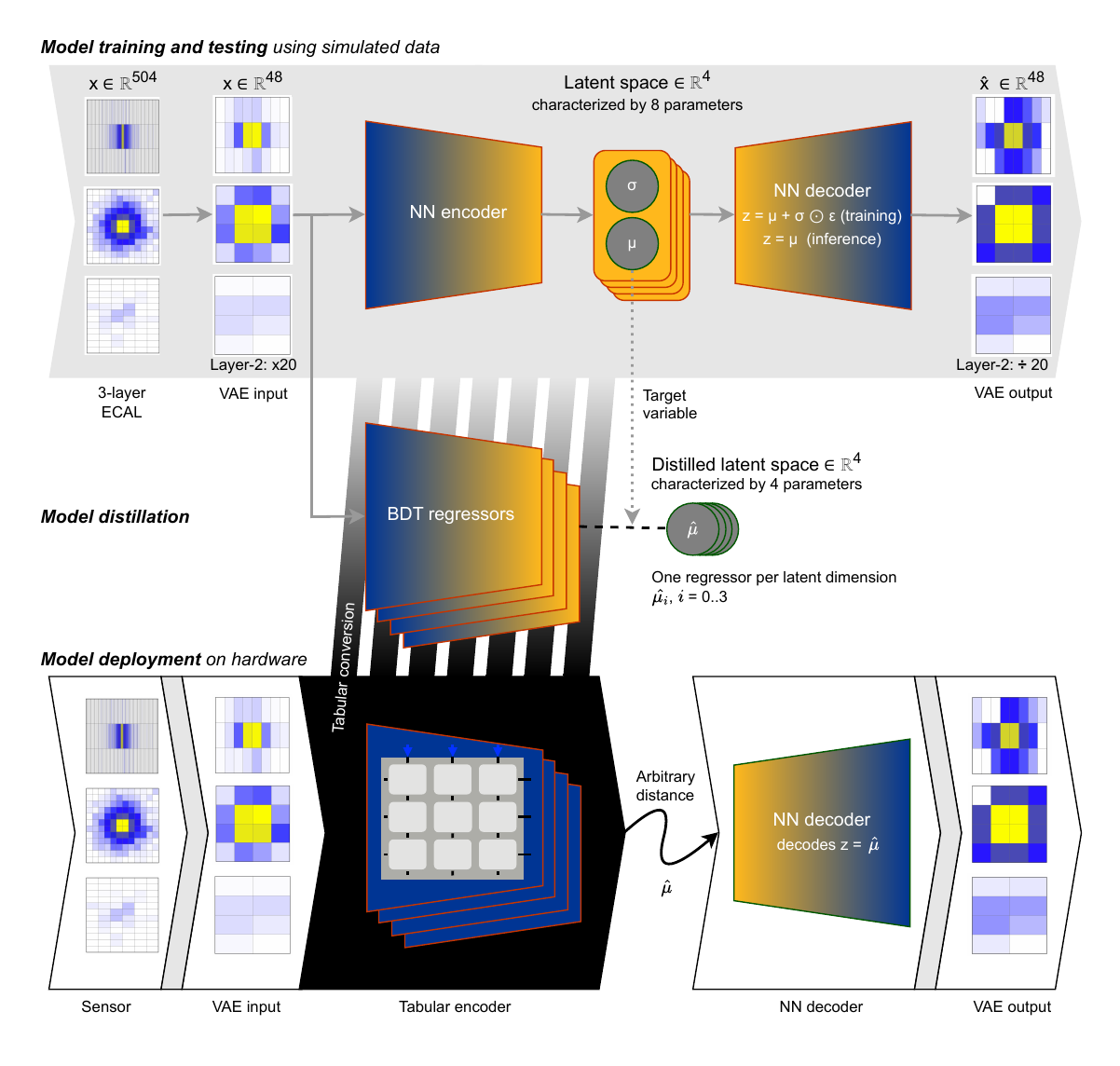}
  \caption{
    Workflow for end-to-end compression on hardware-friendly encoding. Top row: Training and testing shows the encoder mapping $x$ to latent parameters $(\mu,\log\sigma^{2})\in\mathbb{R}^{d}$. Middle row: Distillation uses a set of BDT regressors that emulate the encoder by predicting $\hat{\mu}$ directly from $x$. Latent vectors during inference is deterministic using $z=\mu$. Bottom row: Tabularization flattens the BDT to deploy on the ACAM.
  }
  \label{fig:workflow}
\end{figure}

Model deployment is done in three steps: training the VAE to learn latent representations, distilling the encoder into BDT regressors that emulate its latent mapping, and  deploying the tabularized distilled model on the ACAM. The decoder remains available for offline reconstruction and validation of compressed events. \figurename\:\ref{fig:workflow} visualizes the workflow. 

\begin{figure}[b!]
  \centering
  \includegraphics[width=0.95\textwidth]{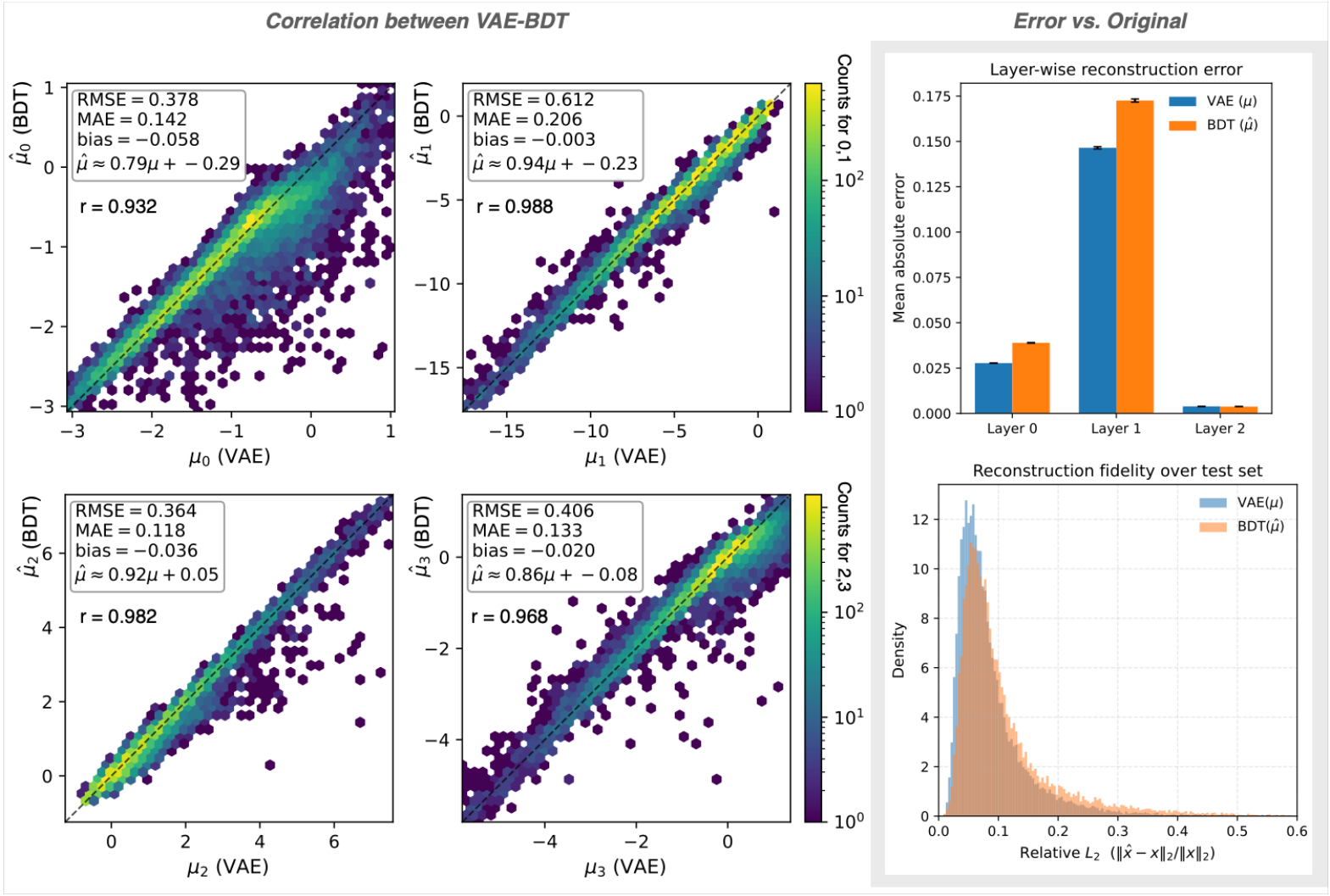}
  \caption{
    Model distillation results. Left group of four plots compares one latent component from the VAE encoder ($\mu_i$) with the corresponding BDT-predicted value ($\hat{\mu}_i$) for incident electrons in the test sample. BDT exhibits strong linearity ($r=0.93$ to $0.99$) and negligible bias, reproducing the encoder outputs with sub-percent mean absolute deviation. Right group of two plots shows quantitative reconstruction fidelity over the test set.  Top-right: Mean absolute error, between VAE and BDT, per ECAL layer; error bars indicate the standard error of the mean across events. Bottom-right: Distribution of the relative $L_{2}$ error over all cells in all layers.
  }
  \label{fig:bdt_scatter}
  \label{fig:reco_errors}
\end{figure}

A set of BDT regressors is trained to mimic the neural encoder to reproduce the latent mean components $\mu_i$ with its estimate $\hat{\mu}_i$. Each component is regressed independently from $\mathbf{x}$, yielding a modular implementation that is naturally parallelizable. Training is performed with TMVA~\cite{TMVA:2007ngy} using statistically independent training and test samples of electron showers. Gradient boosting is used with 200 trees of maximum depth 4, shrinkage factor of 0.05, and stochastic bagging with a 50\% subsample fraction, providing a practical balance between regression accuracy and implementation complexity. The distillation fidelity is evaluated by comparing the BDT-predicted latent vectors $\hat{\mu}$ with the corresponding VAE encoder outputs $\mu$. Strong correlations are observed across all latent dimensions, with Pearson coefficients in the range $r=0.93$ to $0.99$. The four scatter plots of \figurename\:\ref{fig:bdt_scatter} visualizes the correlations. The residuals are narrow and unbiased, indicating that the regressors reproduce the encoder outputs within a small deviation relative to the intrinsic spread of the learned latent space. Additional studies are done and the plots corresponding to the following discussion are given in the \textit{Supplementary Material}. Correlation-matrix studies confirm that the regression uncertainty remains well below the intrinsic variance of the VAE latent space. Residual maps for representative events further show that the differences between $\mu$ and $\hat{\mu}$ reconstructions are small and structureless, with residual distributions centered around zero. This indicates that the distillation preserves both transverse and longitudinal shower features in the compressed representation.

The impact of distillation on the shower reconstruction is evaluated by comparing three representations for each event: the rebinned calorimeter input $\mathbf{x}$, the reconstruction obtained from $\mu$ (VAE in the plots), and the reconstruction obtained from $\hat{\mu}$ (BDT in the plots). This comparison isolates the effect of replacing the encoder, since the same decoder is used for both $\mu$ and $\hat{\mu}$. Representative event displays for the three cases for one simulated electron are shown in the right three columns of \figurename\:\ref{fig:shower_image}. Across the ECAL layers, both reconstructions reproduce the dominant layer-wise energy patterns, including the localized core in Layer~0 and the broader profiles in Layers~1 and 2. Quantitative assessments are made using two complementary metrics: mean absolute error (MAE) per layer and $L_2$ norm per event. The plots in the rightmost column of \figurename\:\ref{fig:reco_errors} shows the results. MAE is computed per ECAL layer, $\sum_{1}^c|x_i-\hat{x}_i|/c$, where $i$ is the cell index for $c$ cells in the considered layer. It is largest in Layer~1, which contains the bulk of the shower energy and therefore dominates absolute deviations, while Layers~0 and~2 exhibit smaller absolute errors. The BDT reconstruction closely tracks the VAE for all layers, with only a modest increase that is most visible in Layer~1. The event-wise relative $L_2$ norm is the square-root of $\sum_1^C(x_i-\hat{x}_i)^2\big/\sum_1^C(x_i)^2$, where $i$ is the cell index for $C$ cells over all layers.
It shows substantial overlap between the two distributions with only a small shift toward larger values for BDT, which is a slight degradation for the approximation.

The BDT is tabularized into hardware-executable JSON configurations using the \texttt{fwXmachina} toolkit; see \emph{Code Availability}. Decision thresholds and leaf outputs are quantized independently into fixed-point integer representations, whose precision is selected to balance model fidelity with hardware constraints. Physics fidelity depends primarily on the precision of the leaf outputs, while the threshold precision plays a secondary role as configurations with 16-bit outputs and 4-bit thresholds preserve the reconstruction quality within statistical uncertainties. In contrast, reducing the output precision below 16 bits leads to measurable degradation. Therefore, we adopt the 16-4 combination for deployment.

\subsection*{ACAM circuit design}

The 6-transistor, 2-memristor (6T2M) ACAM cell  \cite{Li2020ACAM} implements an \emph{interval match} $X \in [L,U]$ on an analog search key $X$ driven on a column as shown previously in the bottom row of \figurename\:\ref{fig:io}. We present, for the first time, an improved version of the cell that exponentially increases the number of levels---or, equivalently, linearly increases the number of bits---stored in a single cell, while linearly increasing the cell size.

\begin{figure}[btp]
  \centering
  \includegraphics[trim={1.4in 4.7in 1.4in 1.6in}, clip, width=\linewidth]{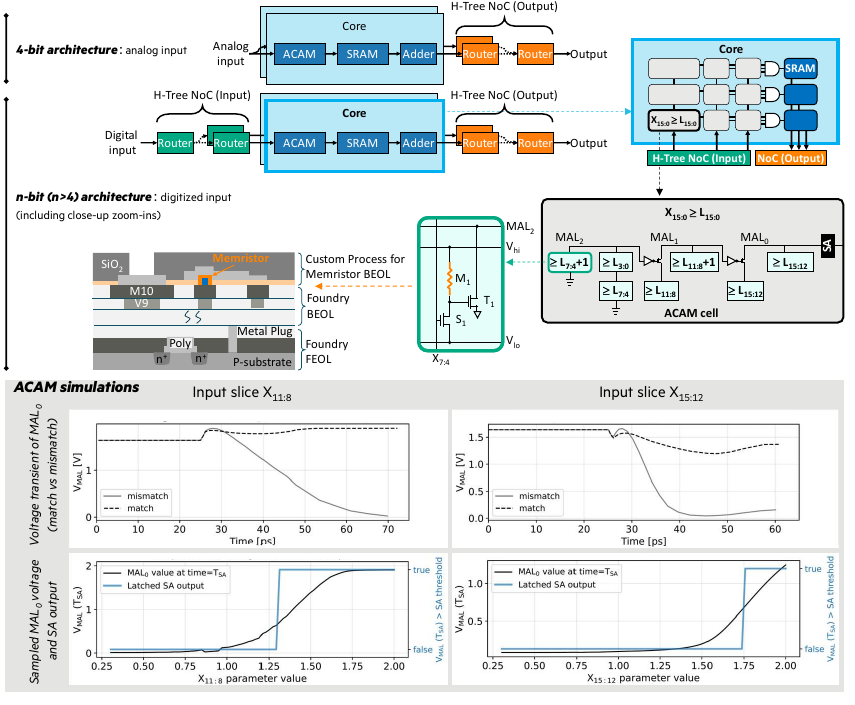}
  \caption{
    ACAM details and results. \textit{4-bit architecture}: Simulation schematic for 4-bit inputs from direct analog sensor data. \textit{n-bit architecture}: Simulation schematic for quantized $n$-bit inputs width for higher precisions (here 16-bit) and successive close-up zoom-ins. Inside the ACAM cell, the MAL$_{0}$ stays charged if, and only if, $X_{15:0}\ge L_{15:0}$; if all the inequalities in the same row of ACAM cells are satisfied, the data in the associated SRAM---corresponding to the selected leaf of the decision tree---is provided as result of the inference. We additionally show the schematic design of an ACAM based on ReRAM \cite{Li2020ACAM}, and the typical material stack cross-section for a custom BEOL manufacturing process of ReRAM \cite{sheng_lowconductance_2019}. \textit{ACAM simulations}: The top row of plots show the time-domain MAL waveforms for match and mismatch, with left and right panels corresponding to input slices $X_{11:8}$ and $X_{15:12}$; the mismatch induced by the $X_{15:12}$ swept values reflects on the MAL state with a smaller time delay due to the ACAM self-loading of the MAL$_{i}$ and of the pull-down devices, with an $X_{11:8}$-induced mismatch propagating in ${\sim}20\,\mathrm{ps}$. The bottom row of plots show the MAL$_{0}$ at the sense-amplifier sampling instant (black box) and the resulting SA decision relative to a fixed threshold; left sweeps $X_{11:8}$ (other input slices are fixed) and right sweeps $X_{15:12}$, with the threshold crossing set by the programmed memristor conductances.
  }
  \label{fig:ACAM-high-precision}
  \label{fig:ACAM-transient}
  \label{fig:Simulation_diagram}
\end{figure}

The match is realized by circuit discharge through two transistors $T$.
In Figures\:\ref{fig:io} and \ref{fig:ACAM-high-precision}, $T_1$ ($T_2$) is for the lower (upper) threshold $L$ ($U$). In the ACAM array, each row corresponds to one MAL and is precharged. The cell only preserves the charge accumulated on MAL, i.e., \emph{matches}, when $L \le X$ \textsc{and} $X \le U$. Physically, the memristor $M_1$ ($M_2$) stores the conductance that encodes $L$ ($U$). The transistor $S_1$ ($S_2$) forms a voltage divider with $M_1$ ($M_2$). The two remaining unlabeled transistors are used for inversion for $U$. The internal sense node controls the MAL-discharge path in the following manner.
\begin{itemize}
  \setlength\itemsep{0pt}
  \item The lower-bound discharge path, flowing through $T_{1}$, discharges MAL if $X < L$.
  \item The upper-bound discharge path, flowing through $T_{2}$, discharges MAL if $X > U$.
  \item If neither discharge path is asserted, the inequality stored in that cell is satisfied.
  In this case, the MAL is not pulled down from that cell.
  \item If all discharge paths on the same row are not asserted, then all inequalities on a root-to-leaf path are satisfied; therefore, the correspondent SRAM on that row stores the inference result for the searched sample.
\end{itemize}
A special case occurs when the $M_{1}$ ($M_{2}$) is set to its minimum (maximum) conductance value. Setting both $M$ in this manner realizes a wildcard called \textsc{don't care}, which provides a match for any value of $X$. Multi-bit precision follows from programming $M$ to discrete conductance levels, achieving $3$ to $4$ bits of precision for each threshold \cite{yin2020fecam,Li2020ACAM}.

A tabularized decision tree, i.e., a binary structure that has been parallelized into a table format with rows and columns, is programmed on the ACAM. The ACAM is capable, in a single cycle, to evaluate a large number of inequalities as large as the number of cells in an ACAM---to perform inference on tree-based models \cite{pedretti_tree-based_2021,Yin2024FeACAM_DRF,pedretti2024x}. When performing such inference, all columns $j$ of the ACAM perform the comparisons in parallel. Given a feature vector $\mathbf{X}$, with each component occupying a vertical column $j$, the MAL on row $i$ stays charged if, and only if, \begin{equation}
\big(X_{j} \in [L_{ij},\,U_{ij}]\big) \textsc{~or~} \big([L_{ij},U_{ij}] \textrm{~is~}\textsc{don't\:care}\big).
\end{equation}
The one-hot-encoded MAL indexes a local SRAM word storing the leaf’s output of the decision tree. In the case where we have an ensemble of decision trees, as is the case in this paper, multiple ACAM tiles fire concurrently and their one–hot row outputs are accumulated, e.g., current summation or digital counting, to realize majority voting or averaging, that corresponds to the desired design. 

Our circuit design depends on the $n$ number of bits for the input features, specifically $n=4$ vs.\ $n>4$. For $n=4$, we use the circuit schematic of an ACAM   where each column value $X_j$ is directly with its stored threshold \cite{Li2020ACAM}; see \figurename\:\ref{fig:io} and \ref{fig:ACAM-high-precision}. This design is appropriate because the precision to represent distinct thresholds is limited to $3$ to $4$ by the memristor conductance programming and by the read noise \cite{pedretti_tree-based_2021, yin2020fecam}.  For $n>4$, we overcome this limitation by preprocessing the inputs as follows. \textit{Bit slicing} is a standard technique in analog IMC to overcome the limited effective precision of crossbar-based MAC operations \cite{le2022precision,bankman2019rram,sinangil20207}. Because memristive devices, e.g., PCM and ReRAM, exhibit conductance variability and noise, directly storing high-precision weights in a single cell is difficult. Instead, an $n$-bit value is decomposed into slices that are mapped to separate columns (spatial slicing) or applied over multiple cycles (temporal slicing), and the resulting partial sums are digitally reweighted and accumulated according to bit position \cite{bankman2019rram,sinangil20207}. Bit slicing is also essential when using binary memories, e.g., SRAM, for analog IMC: if the weights are stored as digital bits, they must be combined via slicing to realize multi-bit analog dot products \cite{sinangil20207,yang202533}.

For $n>4$, we build on this principle to extend the use of bit slicing from its use in MAC calculations to the inequality solving. In order to support higher-precision thresholds in ACAM inequality solvers, we slice the $n$-bit comparisons into sequences of sub-inequalities, and then recombine their outcomes while accounting for their bit-significance.  We introduce a circuit that decomposes an $n$-bit comparison into sub-inequalities whose count scales linearly with the total bit depth. We assume that the slicing happens after the ADC conversion, since the slicing of an analog value in multiple analog slices still requires the ADC conversion of all the most significant parts. For the 16-bit case $X_{15:0}\ge L_{15:0}$, with the subscript denoting, respectively, the most significant bit and the least significant bit for $X$, the following identity holds:
\begin{equation}
  \begin{array}{ccclc}
    X_{15:0}\ge L_{15:0}
      &=& \big[ X_{15:12}\ge L_{15:12}
      &\textsc{and}&
      \big(X_{15:12}\ge L_{15:12}+1\,\textsc{or}\,{\color{blue}X_{11:0}\ge L_{11:0}}\big) \big] \vspace{5 pt}\\ 
    {\color{blue}X_{11:0}\ge L_{11:0}}
      &=& \big[ X_{11:8}\ge L_{11:8}
      &\textsc{and}&
      \big(X_{11:8}\ge L_{11:8}+1\,\textsc{or}\,{\color{red}X_{7:0}\ge L_{7:0}}\big) \big] \vspace{5 pt}\\
    {\color{red}X_{7:0}\ge L_{7:0}}
      &=& \big[ X_{7:4}\ge L_{7:4}
      &\textsc{and}&
      \big(X_{7:4}\ge L_{7:4}+1\,\textsc{or}\,X_{3:0}\ge L_{3:0}\big) \big]\\
  \end{array}
\end{equation}
These recursive relations reduce $X_{15:0}\ge L_{15:0}$ into an expression composed of four inequalities each with 4-bit precision, i.e., $X_{15:12}\ge L_{15:12}$, $X_{11:8}\ge L_{11:8}$, $X_{7:4}\ge L_{7:4}$, and $X_{3:0}\ge L_{3:0}$. This logic is illustrated in the zoom of the ACAM cell in \figurename\:\ref{fig:ACAM-high-precision}.

Assuming that a 4-bit threshold can be stored in a single ReRAM device, the recursive application of this identity over 4-bit slices yields the high-precision implementation shown in \figurename\:\ref{fig:ACAM-high-precision}. This configuration trades-off the search latency and compactness of the simple 6T2M cell with the capability to extend the number of distinct thresholds, which grows exponentially (the number of bits increases linearly) with the area of the cell. By storing \textsc{don't care} values in a subset of 4-bit cells, it is possible to obtain a configuration suitable for the 12-bit inequality case. Equivalently, the 12-bit inequality can be solved relaxing the design constraints of the memory cells, assuming that each ReRAM stores a 3-bit sliced information. The composition of multiple 16-bit inequalities with standard CMOS logic gates allows to resolve, in parallel and in a single step, multiple inequalities with a number of the thresholds levels that is orders of magnitude higher than the one achievable by standard ACAM cells.

The design and experimental verification is discussed. The 4-bit case is presented in Ref.~\cite{Li2020ACAM}, where it is extensively discussed how the cell simulations inform the larger memory array design, as well as the number of bits per cell. For the higher bits, we verified the 16-bit memory cell behavior through SPICE simulations with the following results. The first row of plots in \figurename\:\ref{fig:ACAM-transient} show the variation in time of the MAL node for the match and mismatch cases for two input slices $X_{11:8}$ and $X_{15:12}$. A mismatch induced by $X_{11:8}$ generates a mismatch with a propagation delay of about $20$\,ps, due to the self-loading of the ACAM for the proposed configuration. The bottom row of plots in the same figure shows the MAL$_0$ voltage at the time the MAL$_0$ is compared against a fixed threshold by the sense amplifier (SA), and the resulting true/false value that the SA provides. The left plot shows the simulated MAL$_0$ values and the SA output sample when the slice $X_{11:8}$ is swept along the entire analog range and the other input slices are fixed. Similarly, in the right plot, the input slice $X_{15:12}$ is swept and the other slices are fixed. The input voltage for which the MAL$_{0}$ overcomes the fixed threshold is determined by the memristor states.

\section*{Data availability}
The electromagnetic calorimeter shower dataset used in this work was obtained from Mendeley Data at \url{https://data.mendeley.com/datasets/pvn3xc3wy5/1} \cite{ecal_mendeley}. All derived datasets produced by our pipeline---e.g., files related to the model (VAE, BDT, test sample) and files related to the variables (reordered regressed latent vectors, reconstructed shower outputs, shower-variables)---can be reproduced from the shower dataset using the code in the \textit{Code availability}.

\section*{Code availability}
Physics studies, model training, and tabularization are done using \href{https://fwx.pitt.edu}{\texttt{\textsc{fwXmachina}}}, \url{https://fwx.pitt.edu}. The branch used in this work is available at \url{https://gitlab.com/PittHongGroup/starcoder-dev}. Hardware studies are carried out using X-TIME: a simulator framework for tree-based ML accelerators based on analog content-addressable memories available at \url{https://github.com/HewlettPackard/X-TIME} and the SST Structural Simulation Toolkit at \url{https://sst-simulator.org/}.

\section*{Acknowledgments}

We thank Philip Chang for discussions regarding the $\mu$C.
Work performed in Dietrich School Electronics Shop Core Facility (RRID:SCR 025113) and services and instruments used in this project were graciously supported, in part, by the University of Pittsburgh. 
YE is supported by the National Science Foundation (award no.\ PHY-1948993).
TMH and RG are supported by the US Department of Energy (award no.\ DE-SC0007914).

\section*{Author contribution statement}

RG designed the ML algorithm, created the derived data sample, and performed the physics studies.
STR executed the model tabularization.
JM implemented the circuit design on the SST.
AN worked on the layout and SPICE simulations.
YE did the FPGA implementation.
LB coordinated the overall effort while TMH and JI took a supervisory role.
TMH compiled the manuscript from contributions by RG, JM, LB, STR, and YE.
All authors reviewed the manuscript. 

\section*{Competing interests statement}

The authors declare competing interests. TMH and STR are part of a patent on the firmware design of the autoencoder with the University of Pittsburgh as US Patent Application Publication No.\ US 2024/0054399.
Other authors declare no competing interests.

\input{supplementary_arxiv}

\end{document}

%% file: supplementary_arxiv.tex
\newpage
\section*{Supplementary Material}

\setcounter{figure}{0}
\setcounter{table}{0}
\renewcommand{\figurename}{Supplementary Figure}
\renewcommand{\tablename}{Supplementary Table}

\begin{table*}[hbp!]
  \centering
  \caption{
    One-dimensional shower observables used to assess the quality of reconstructed electromagnetic calorimeter showers. Layer index $\ell\in\{0,1,2\}$.
    }
  \label{tab:shower_observables_1d}
  \renewcommand{\arraystretch}{1.35}
  {\small
  \begin{tabular}{p{0.19\textwidth} p{0.34\textwidth} p{0.36\textwidth}}
    \toprule
    \textbf{Shower variable} & \textbf{Formula} & \textbf{Notes} \\
    \midrule
    Layer energy & $E_\ell=\sum_{\text{c}} E_c$ & Energy in cell $c$ per layer $\ell$ (GeV) \\
    Total energy & $E_{\mathrm{tot}}=\sum_\ell E_\ell$ & Over all layers (GeV) \\
    Energy fraction & $f_\ell=E_\ell\big/E_{\mathrm{tot}}$ & Fraction in layer $\ell$ (unitless) \\
    Shower depth & $s_d=\sum_\ell E_\ell\,\ell\big/E_{\mathrm{tot}}$ & Energy-weighted depth ($\ell$ units) \\
    Shower depth width & $\sigma_{s_d}=\bigg(\dfrac{\sum_\ell E_\ell\,\ell^2}{E_{\mathrm{tot}}}-\bigg(\dfrac{\sum_\ell E_\ell\,\ell}{E_{\mathrm{tot}}}\bigg)^2\bigg)^{1/2}$ & Spread of the depth distribution ($\ell$ units) \\
    Layer lateral width & $\sigma_\ell=\bigg(\dfrac{\sum_c E_c\,\eta_c^2}{E_\ell} - \bigg(\dfrac{\sum_c E_c\,\eta_c}{E_\ell}\bigg)^2\bigg)^{1/2}$ & Weighted width along $\eta$ in layer $\ell$ (mm) \\
  \bottomrule
  \end{tabular}
  }
\end{table*}

\FloatBarrier

\begin{figure}[htb!]
  \centering
  \includegraphics[width=\textwidth]{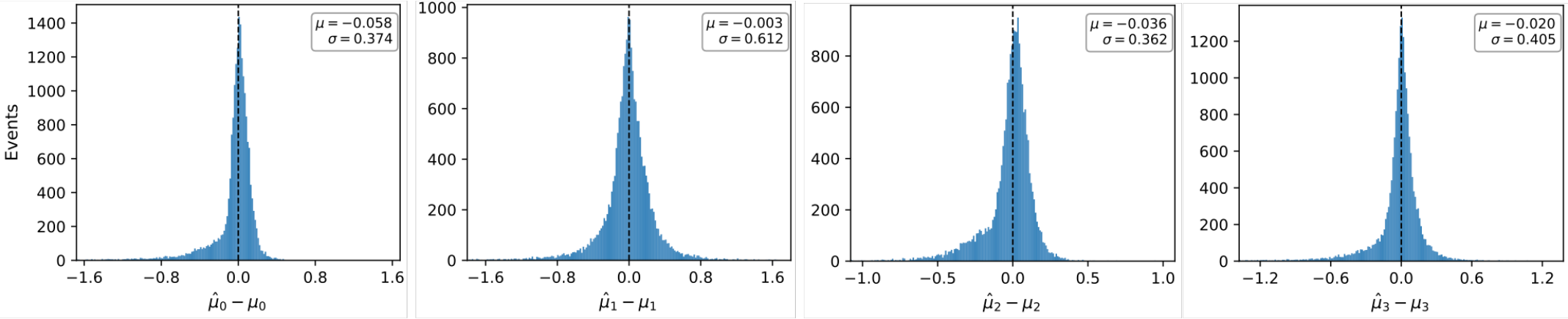}
  \caption{
    Residual distributions of the regressed latent variables for the scenario presented in the paper. Shown are the per-component differences between VAE and BDT. The vertical dashed line is at zero.
  }
  \label{fig:bdt_residual}
\end{figure}

\begin{figure}[htb!]
  \centering
  \includegraphics[width=0.55\textwidth]{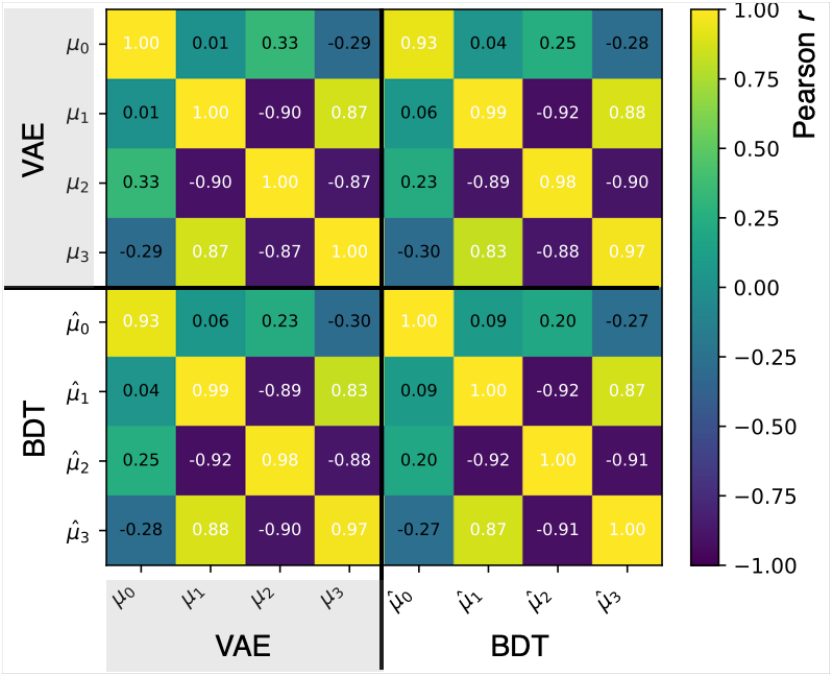}
  \caption{
    Correlation matrix between the VAE and BDT for the scenario presented in the paper. Pearson correlation coefficients are shown in each box. The diagonal values for VAE-BDT are all above $0.9$; the off-diagonal values for VAE-BDT are similar to the corresponding values for VAE-VAE and BDT-BDT.
  }
  \label{fig:bdt_correlation}
\end{figure}

\begin{figure}[p!]
  \centering
  \includegraphics[width=1.0\textwidth]{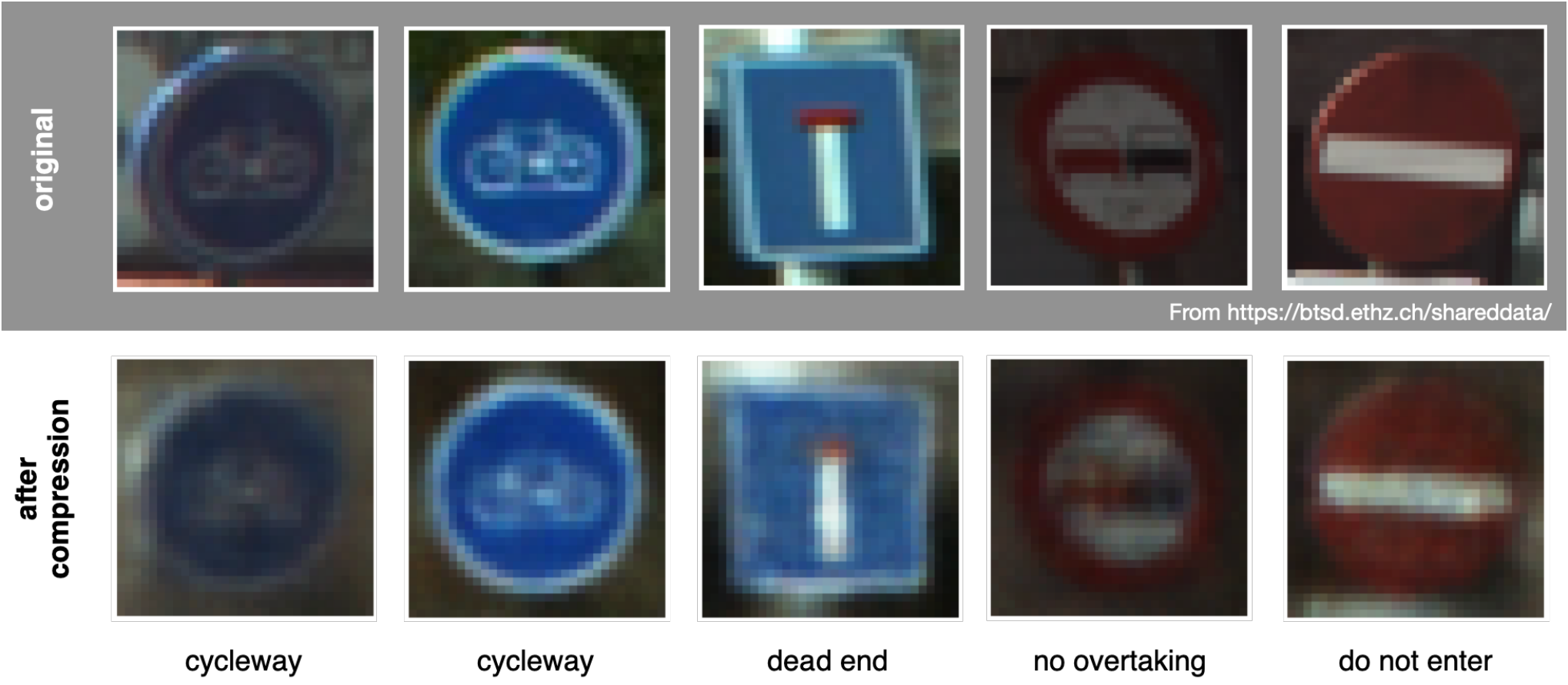}
  \caption{
    Compression and distillation example on color photos of Belgian traffic signs. The images show the same workflow used in this paper. Top row (original images): $32\times32$ RGB cropped sign images from the BelgiumTS/BTSC dataset~\cite{BelgiumTS_dataset,Mathias2013IJCNN}. Bottom row (images after compression-decompression): reconstructions obtained by feeding the BDT-predicted latent vectors $\hat{\mu}$ into the VAE decoder. Details: \emph{Model} used is a convolutional encoder with three $4{\times}4$ \textsc{Conv} layers, each using stride $2$ (with \textsc{ReLU} activations and $3$, $32$, $64$, and $128$ feature maps), which maps an input image to a latent representation; this produces the $\mu$ and $\log\sigma^2$ parameters of a $d{=}32$ dimensional Gaussian latent space. During training a latent code is sampled via the standard reparameterization $z{=}\mu{+}\sigma{\cdot}\epsilon$; a symmetric transposed-convolution decoder (three $4{\times}4$ \textsc{ConvTranspose} layers with \textsc{ReLU}, followed by a Sigmoid output layer) reconstructs the image. This corresponds to a compression of a $32{\times}32{\times}3$ (3072 pixel intensities) to a 32d latent vector, i.e., $96$x reduction. \emph{Distillation} replaces the neural-network encoder with a BDT regressor to predict the latent space variables of the VAE (here, the latent means $\mu$).
  }
  \label{fig:signs}
\end{figure}

\renewcommand\refname{Supplementary References}